\documentclass[11pt,a4paper]{amsart}
\pdfoutput=1

\usepackage{amsthm,amssymb,amsmath,epic,eepic,float,subfigure}
\usepackage{graphics,graphicx,geometry}

\newcommand{\neel}{{\rm N\acute{e}el}}

\def\leq{\leqslant}

\newcommand{\cC}{\mathcal C}

\newcommand{\cH}{\mathcal H}
\newcommand{\cI}{\mathcal I}

\newcommand{\cM}{\mathcal M}
\newcommand{\cN}{\mathcal N}
\newcommand{\cO}{\mathcal O}

\def\ll{    \left\lgroup \!}
\def\rr{\! \right\rgroup   }

\def\det{\operatorname{det}}

\begin{document}
\title[Partial N\'eel states and on-shell Bethe states]
{Overlaps of Partial N\'eel States and Bethe States}
\author{O. Foda}  
\address{
School of Mathematics and Statistics,
The University of Melbourne,
Parkville, Victoria 3010, Australia
}
\email{omar.foda@unimelb.edu.au}

\author{K. Zarembo}
\address{
Nordita, KTH Royal Institute of Technology and Stockholm University,
Roslagstullsbacken 23, SE-106 91 Stockholm, Sweden
{\tt and}
Department of Physics and Astronomy, Uppsala University
SE-751 08 Uppsala, Sweden
}

\email{zarembo@nordita.org}
\begin{abstract}
Partial N\'eel states are generalizations of the ordinary N\'eel (classical anti-ferro\-mag\-net) state that 
can have arbitrary integer spin. We study overlaps of these states with Bethe states. We first identify 
this overlap with a partial version of reflecting-boundary domain-wall partition function, and then derive 
various determinant representations for off-shell and on-shell Bethe states.
\end{abstract}
\begin{flushright}
\footnotesize
%\texttt{ITEP-TH-nn/yy}\\
\texttt{NORDITA-2015-133} \\
\texttt{UUITP-27/15}
\vspace{0.6cm}
\end{flushright}

\maketitle

%SEC.01
\section{Introduction}
\label{section.01}

\subsection{Overview}
The N\'eel state is the simplest state with antiferromagnetic ordering, differing though from the true ground 
state of the antiferromagnetic XXX spin-$\frac12$ chain, which is more complicated. One may ask how close the 
N\'eel state is to the true eigenstate of the spin-chain Hamiltonian. The answer to this question is actually 
known, as the overlap of the N\'eel state with any given eigenstate can be explicitly calculated. As pointed 
out in \cite{Pozsgay:2009}, the overlap is related  to the partition function of the six-vertex model on 
a rectangular lattice with reflecting boundary conditions. A determinant representation for this overlap was 
obtained by Tsuchiya \cite{Tsuchiya:qf}. 
Restricting Tsuchiya's expression to Bethe eigenstates requires an extra step, and leads to a simpler, more 
compact determinant expression  \cite{Brockmann:2014a, Brockmann:2014b, Brockmann:2014c}. Applications of 
these results range from condensed-matter physics \cite{Brockmann:2014a, Brockmann:2014b, Brockmann:2014c, 
Brockmann:2014d, Mazza:2015} to string theory \cite{deLeeuw:2015hxa} and algebraic combinatorics 
\cite{kuperberg2002symmetry}.

In this note, we generalize Tsuchiya's result to the case of partition functions of six-vertex model configurations 
with domain-wall boundary conditions that are only partially reflecting. These partially reflecting domain-wall 
boundary conditions are related to Tsuchiya's \cite{Tsuchiya:qf} in the same way 
that the partial domain-wall boundary conditions and partition function in \cite{Foda:2012yg} are related to Korepin's 
domain-wall boundary conditions \cite{Korepin:1982gg}, and Izergin's determinant expression for the corresponding partition 
function \cite{Izergin1987}. The on-shell version of the partition function that we study describes overlaps of the Bethe 
states with partial, or generalized N\'eel states \cite{Brockmann:2014c, deLeeuw:2015hxa}. Our derivations closely follow 
those in \cite{Pozsgay:2009, Tsuchiya:qf, Brockmann:2014a}.

\subsection{Outline of content}
In section {\bf \ref{section.02}}, we recall basic definitions related to the XXX spin-$\frac12$ chain and the rational 
six-vertex model. 
In {\bf \ref{section.03}}, we recall the definition of Tsuchiya's reflecting-boundary boundary conditions, and 
the corresponding partition function, then introduce partial versions thereof, which we subsequently calculate.
In {\bf \ref{section.04}}, we compute the overlap of a partial N\'eel state and a parity-invariant highest-weight 
on-shell Bethe state.
In {\bf \ref{section.05}}, we collect a number of remarks.
and in appendix {\bf A}, we explain the reduction of the $[ M \! \times \! M ]$ determinant partition function 
to an $[ \frac{M}{2} \! \times \! \frac{M}{2} ]$ one.

%SECTION.02
\section{The XXX spin-$\frac12$ chain and the rational six-vertex model}
\label{section.02}

We restrict our attention to the XXX spin-$\frac12$ 
chain of length $L=2N$ \cite{Baxter:1972hz}. Each site carries an up-spin $\uparrow$, or 
a down-spin $\downarrow$. The Hamiltonian $\cH$ acts on $\ll {\cC}^2 \rr^{\otimes L}$ as

\begin{equation}\label{HHam}
 \cH = \sum_{l=1}^{2N} \ll 1-P_{l,l+1} \rr,
\end{equation}

\noindent where $P_{l, l+1}$ permutes the spins on two adjacent sites labeled $l$ and $l+1$.

\subsection{The $R$-matrix and the Yang-Baxter equation}

%FIGURE.01
\begin{figure}[t]
\begin{center}
\centerline{\includegraphics[width=10cm]{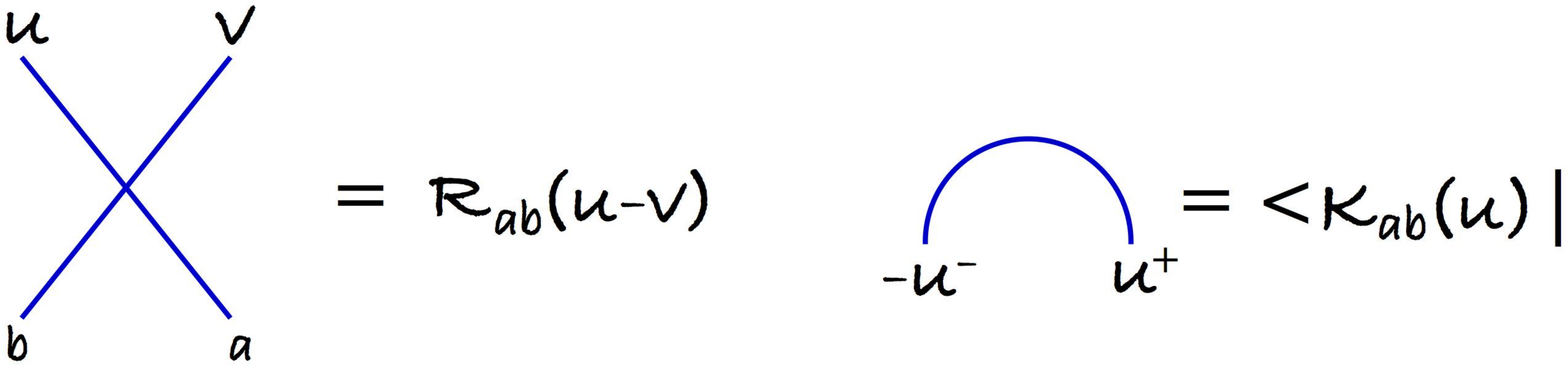}}
\caption{
{\it The $R$-matrix and the boundary state.}
}
\label{RK}
\end{center}
\end{figure}

%FIGURE.02
\begin{figure}[t]
\begin{center}
 \centerline{\includegraphics[width=6cm]{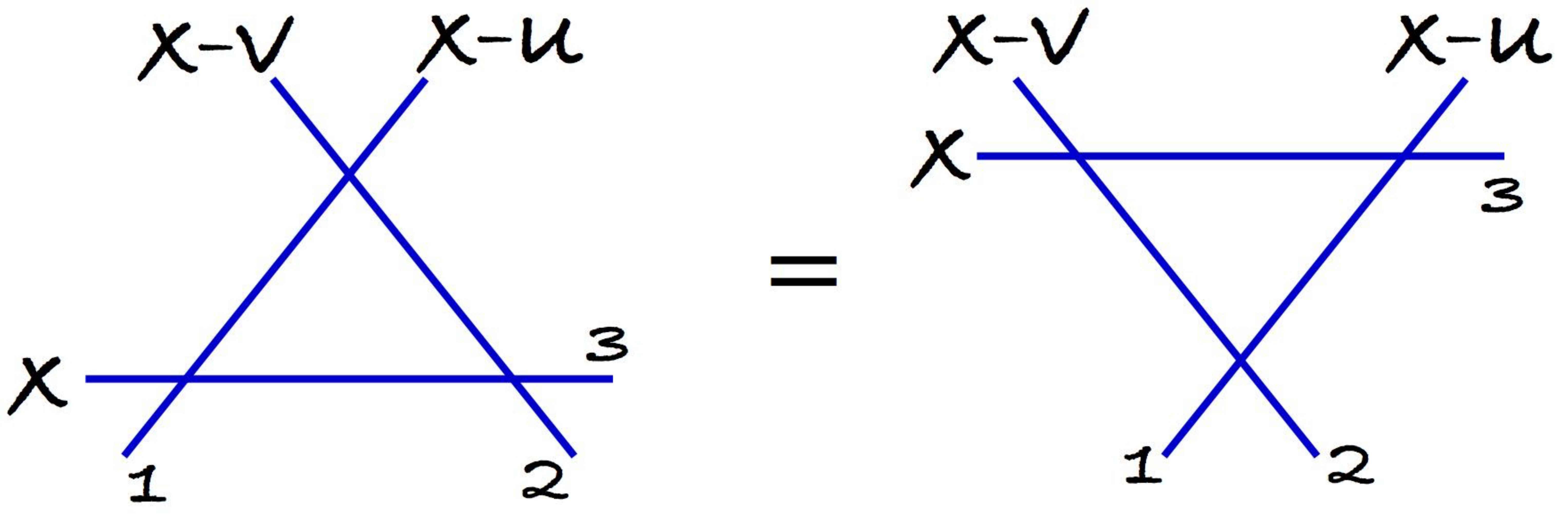}}
\caption{
{\it The Yang-Baxter equation.}
}
\label{YBfig}
\end{center}
\end{figure}

The key object in the Bethe Ansatz solution of the XXX spin-$\frac12$ chain is the $R$-matrix \cite{inverse}, 
represented in Figure {\bf \ref{RK}}. The $R$-matrix, $R_{ab} (u)$, acts on the tensor product 
of two spins labelled $a$ and $b$, and depends on a complex spectral parameter $u$,

\begin{equation}
 R_{ab}(u) = u + iP_{ab},
\end{equation}

\noindent where $P_{ab}$ is the permutation operator. Most importantly, the $R$-matrix satisfies the 
Yang-Baxter equation, 

\begin{equation}
 R_{12}(u-v) R_{13}(u) R_{23}(v) = R_{23}(v) R_{13}(u) R_{12}(u-v),
\end{equation}

\noindent represented in Figure {\bf \ref{YBfig}}. 

\subsection{Notation and conventions.}
We will use the shorthand notation 

\begin{equation}
y^\pm=y\pm\frac{i}{2}\,.
\end{equation}

\noindent The $R$-matrix acts from south to north or, for the vertical crossing, from south-east to north-west, 
as represented in Figure {\bf \ref{RK}}. If each line carries a rapidity variable, the argument of the $R$-matrix 
is the difference of the two rapidity variables.

\subsection{The $B$-operator}
The states of the XXX spin-$\frac12$ chain are generated by the $B$-ope\-ra\-tors. These operators are 
constructed by multiplying the $R$-matrices along the spin-chain threaded with a single auxiliary space,

\begin{equation}
\label{B.operator.unpaired}
 B_{\mathbf{y}}(x)=
 \left\langle \uparrow_a\right|
 R_{1,     a}\left(x+y_1\right)
 R_{2,     a}\left(x-y_2\right) 
 \cdots 
 R_{L-1,  a}\left(x+y_{L-1}\right)
 R_{L,    a}\left(x-y_{L  }\right)
 \left| \downarrow_a\right\rangle.
\end{equation}

\subsection{Inhomogeneity variables}
The variables $y_k$, $k=1, \cdots, L$, are the quantum-space or inhomogeneity variables.
They play an important r\^ole in the intermediate steps of the derivations, but they are not necessary for 
diagonalizing the Hamiltonian (\ref{HHam}).
The inhomogeneity variables are common to all $B$-operators used to construct the state. In this note, 
we do not consider the most general inhomogeneity variables as in (\ref{B.operator.unpaired}), but focus on 
parity-invariant Bethe states such that the inhomogeneity variables are paired, as in 

\begin{equation}
\label{B-op}
 B_{\mathbf{y}}(x) =
 \left\langle \uparrow_a \right|
 R_{1, a}\left(x+y_1^-\right)
 R_{2, a}\left(x-y_1^+\right) 
 \cdots 
 R_{2N-1,a} \left(x+y_N^-\right)
 R_{2N,  a} \left(x-y_N^+\right)
 \left| \downarrow_a\right\rangle.
\end{equation}

\noindent The restriction to paired inhomogeneity variables will be important later on, when we consider
overlaps of Bethe states and the boundary states introduced in section {\bf \ref{boundary.state}}. 

\subsection{On-shell Bethe states}
The Bethe states of the XXX spin-$\frac12$ chain are constructed by applying the $B$-operators 
on the ferromagnetic vacuum, $\left|0\right\rangle=\left|\uparrow \ldots \uparrow \right\rangle$, 
of the spin-chain, 

\begin{equation}\label{Bethe-state}
 \left|\mathbf{x}\right\rangle=B(x_M)\ldots B(x_1)\left|0\right\rangle, 
\end{equation}

\noindent 
where $B(x)=B_{\mathbf{0}}(x)$, that is, the $B$-operator with all $y_a$-variables set to zero.
For a Bethe state to be an eigenstate of the Hamiltonian, the rapidity variables $x_j$ must satisfy the Bethe 
equations,

\begin{equation}
\label{bethe.equations}
 \ll \frac{x_j^+}{x_j^-} \rr^{2N}=
 \prod_{k\neq j} \frac{x^+_j-x^-_k}{x^-_j-x^+_k}\,.
\end{equation}

\noindent As usual, we call Bethe states with rapidity variables subject to the Bethe equations on-shell  
states. Generic Bethe states, that are not eigenstates of the Hamiltonian, are referred to as off-shell states. 

\subsection{Partial N\'eel states}
\label{partial.neel.state}

Given the definition of the N\'eel state, on an XXX spin-$\frac12$ chain of length $L=2N$,  

\begin{equation}
\left| \neel \right\rangle = 
\left| \uparrow \downarrow \uparrow \downarrow \cdots \uparrow \downarrow \right\rangle+
\left|\downarrow \uparrow \downarrow \uparrow \cdots \downarrow \uparrow \right\rangle,
\end{equation}

\noindent and an integer $M$, such that $0 \leq M \leq N$, we define an $M$-partial N\'eel state as 
\cite{Brockmann:2014c,deLeeuw:2015hxa}

\begin{equation}\label{generalized-Neel}
\left| \neel_M \right\rangle = 
\sum_{
{l_1 < \cdots < l_M}
\atop
{|l_i-l_j| = 0 \, \textit{mod} \, 2}
}
\left| \cdots \uparrow \downarrow_{l_1} \uparrow \cdots 
              \uparrow \downarrow_{l_2} \uparrow \cdots 
	      \uparrow \downarrow_{l_M} \uparrow \cdots
\right\rangle.
\end{equation}

\noindent For $M=N$, the state has an equal number of up- and down-spins, and we recover the original 
N\'eel state, $\left| \neel_N \right \rangle=\left| \neel \right\rangle$. 
 
\subsection{Matrix product states}
 
The matrix product state, \textit{MPS}, is defined as

\begin{equation}\label{MPS-state}
 \left| \textit{MPS\,} \right\rangle =
 \mathop{\mathrm{tr}}\nolimits_a 
 \ll
 t_a^\uparrow \left|\uparrow_1\right\rangle + t_a^\downarrow 
 \left|\downarrow_1 \right\rangle
 \rr
 \otimes
 \cdots 
 \otimes
 \ll
 t_a^\uparrow \left| \uparrow_L \right\rangle + t_a^\downarrow \left|\downarrow_L \right \rangle
 \rr,
\end{equation}

\noindent where $t^\uparrow =\sigma _1$ and $t^\downarrow =\sigma _2$. Following \cite{deLeeuw:2015hxa}, 
all partial N\'eel states can be obtained from \textit{MPS} by projecting on subspaces with a definite 
number of up- and down-spins. Denoting the projector on the state with $M$ down-spins by $P_M$, we have,

\begin{equation}\label{MPS<->Neel}
 \left|\neel_M\right\rangle = 2^{L} 
 \ll \frac{i}{2} \rr^M
 P_M \left| \textit{MPS\,} \right\rangle + S^-\left| \Psi \right\rangle,
\end{equation}

\noindent where $S^-$ is the spin-lowering operator. The last term does not contribute to the overlap with 
a Bethe state that is annihilated by $S^+$, such as the highest-weight on-shell Bethe states of the homogeneous 
XXX spin-$\frac12$ chain.
 
\subsection{The boundary state}
\label{boundary.state}
As noted in \cite{Pozsgay:2009}, the N\'eel state can be constructed from the boundary state 
associated with the diagonal reflection matrix. The overlap of a N\'eel state on a one-dimensional lattice
of length $L=2N$ and a Bethe state that is not necessarily on-shell, characterised by $N$ rapidity variables, 
is equal to the partition function of the six-vertex model on a rectangular lattice that has $N$ horizontal 
lines and $2N$ vertical lines, with reflecting-boundary domain-wall boundary conditions. 
Following \cite{Tsuchiya:qf}, the latter is an $[ N \! \times \! N ]$ determinant \cite{Tsuchiya:qf}. 
We extend this construction by effectively allowing for non-diagonal scattering off the boundary. The latter 
boundary state reduces to a partial N\'eel state (\ref{generalized-Neel}) for appropriate values of the variables. 
The boundary state is defined as

\begin{equation}\label{boundary-state}
 \left\langle K_{ab}(u)\right|=
  \left\langle \uparrow \downarrow \right| \ll u^++\xi \rr
 +\left\langle \downarrow \uparrow \right| \ll u^+-\xi \rr
 +\left\langle \uparrow \uparrow \right|\lambda u^+.
\end{equation}

\noindent Although the boundary state is associated with two lattice lines, it depends on a single rapidity 
$u$. The alternating inhomogeneity variables in (\ref{B-op}) lead to one independent inhomogeneity variable 
for each boundary state. The variables $\xi $ and $\lambda $ are arbitrary, but fixed complex numbers. 

The boundary state is the cross-channel representation of the reflection matrix \cite{Cherednik:1985vs,Sklyanin:1988yz}. 
The most commonly used boundary state is the neutral one with $\lambda =0$, which corresponds to the diagonal 
reflection \cite{Cherednik:1985vs,Sklyanin:1988yz}. It is this diagonal reflection matrix that was used in the 
derivation of \cite{Tsuchiya:qf}. We extend this result by adding the last, two-spins-up term
\footnote{\, 
One can, in principle, also add a term with two down-spins that has a weight $\mu u^+$, which would correspond 
to the most general rational solution of the reflection equation \cite{deVega:1993xi}. It would be interesting 
to generalize the computation of the overlaps and the partition functions defined below to this case as well. 
}, allowed by the consistency conditions for integrable boundary scattering \cite{deVega:1993xi}. When representing 
the boundary state in terms of a diagram, we assume that it can only connect spins whose rapidity variables add 
to $i$, as shown in Figure {\bf \ref{RK}}. 

%FIGURE.03
\begin{figure}[t]
\begin{center}
\centerline{\includegraphics[width=9cm]{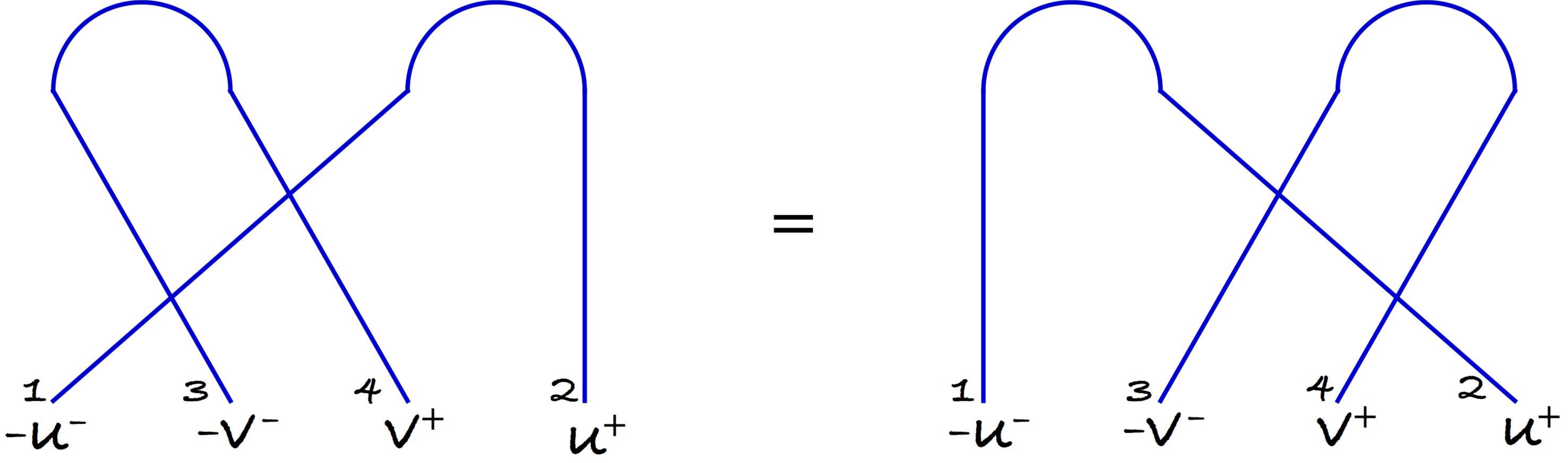}}
\caption{
{\it The reflection equation for the boundary state.}
}
\label{Reflectionfig}
\end{center}
\end{figure}

\subsection{The reflection equation}
The boundary state obeys the reflection equation \cite{Cherednik:1985vs,Sklyanin:1988yz}, see Figure 
{\bf \ref{Reflectionfig}}, which, in our notation, takes the form

\begin{equation}
\label{reflection-eq}
\left\langle K_{23}(v)\otimes K_{12}(u)\right|R_{14}(u+v)R_{13}(u-v)
=
\left\langle K_{12}(u)\otimes K_{34}(v)\right|R_{23}(u+v)R_{24}(u-v).
\end{equation}

\subsection{The overlap of a partial N\'eel state and a parity-invariant highest-weight on-shell Bethe 
state}
The object of our interest is the overlap

\begin{equation}\label{genTsoverlap}
 Z_{\mathbf{x}\mathbf{y}} =
 \left\langle K_{12} \left(y_1\right) 
 \otimes
 \cdots 
 \otimes 
 K_{2N-1,2N}
 \left(y_N \right)\right|
 B_{\mathbf{y}}(x_M) \cdots B_{\mathbf{y}} \left(x_1\right) \left| 0 \right\rangle.
\end{equation}

\noindent It depends on two sets of rapidity variables 
$\{x_j\}_{j=1, \cdots, M}$, and
$\{y_a\}_{a=1, \cdots, N}$. We will not put further constraints on these variables at this point. 
The spectral parameters of the boundary state, as in Figure {\bf \ref{RK}}, are correlated with 
the arguments of the $R$-matrices in (\ref{B-op}). This is the reason for pairing them, instead 
of keeping them arbitrary.

The overlap of the on-shell Bethe states of the homogeneous spin-chain with the partial N\'eel 
states (\ref{generalized-Neel}) can be obtained by setting $y_a=0$, $\lambda =- 2i$, and $\xi $ 
to $\pm \frac{i}{2}$, and imposing the Bethe equations on the rapidity variables $x_j$,

\begin{equation}\label{six-vertex-to-Neel}
 \langle \neel_M | \mathbf{x} \rangle
 = 
 \ll -i \rr^M 
 \ll
 Z_{\mathbf{x}\, \mathbf{0}} | {}_{\lambda =-2i,\,\xi = \frac{i}{2}}
 +
 Z_{\mathbf{x}\, \mathbf{0}} | {}_{\lambda =-2i,\,\xi =-\frac{i}{2}}
\rr.
\end{equation}

\noindent This follows from the structure of the boundary state (\ref{boundary-state}). Setting 
$\xi =\pm \frac{i}{2}$ leaves only two terms in the boundary state (\ref{boundary-state}) whose $N$-th 
tensor power then generates the sum of all partial N\'eel states. Since any given Bethe state has exactly 
$M$ down-spins, only the $M$-th N\'eel state can have a non-zero overlap with it.

%FIGURE.04
\begin{figure}[t]
\begin{center}
 \centerline{\includegraphics[width=7cm]{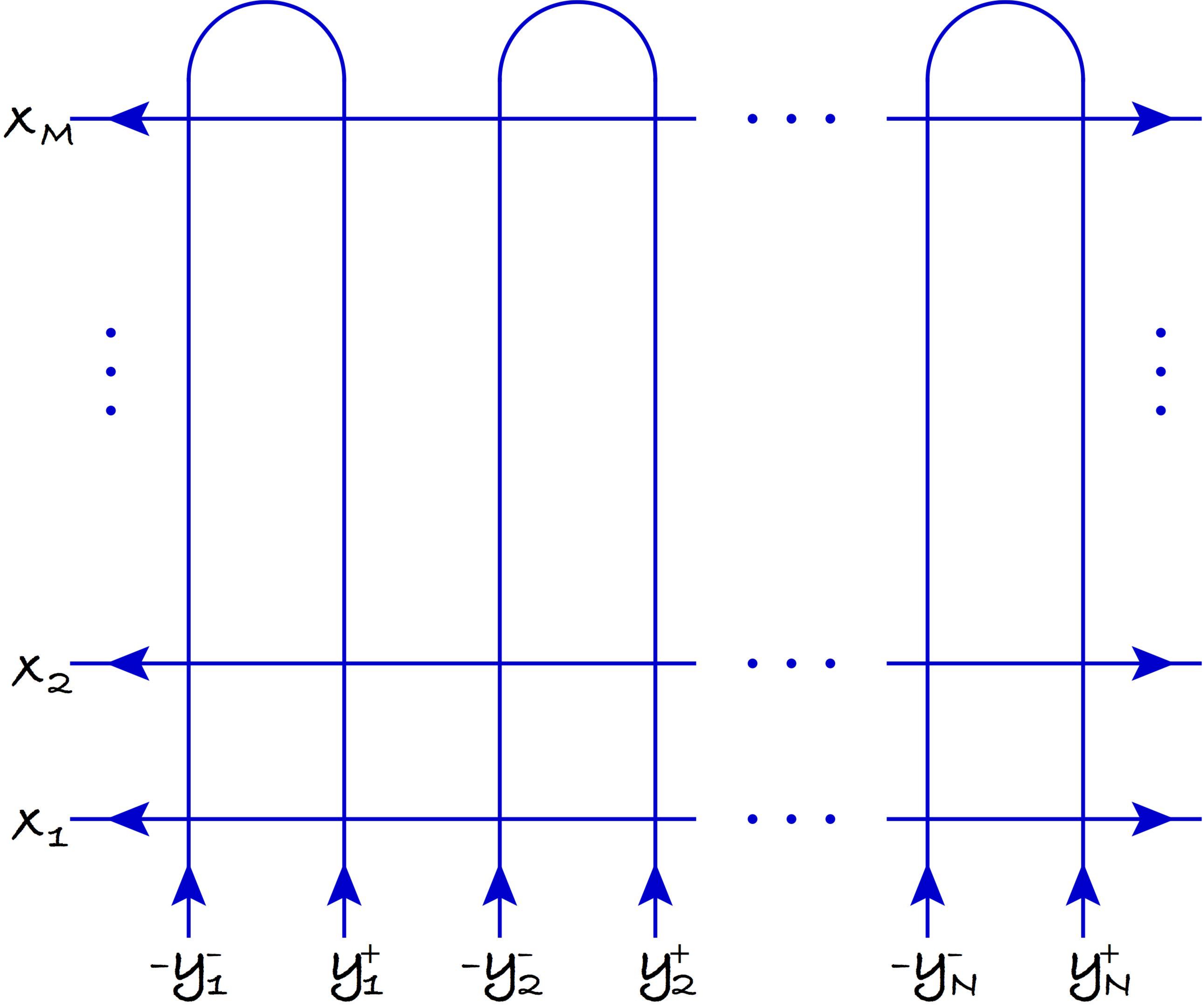}}
\caption{
{\it The partition function of the six-vertex model. Each link of the lattice carries an up- or a down-spin.  
Summation over all spin variables is implied unless the direction of the spin is explicitly indicated.} 
}
\label{Zxy}
\end{center}
\end{figure}

%FIGURE.05
\begin{figure}[t]
\begin{center}
\centerline{\includegraphics[width=13cm]{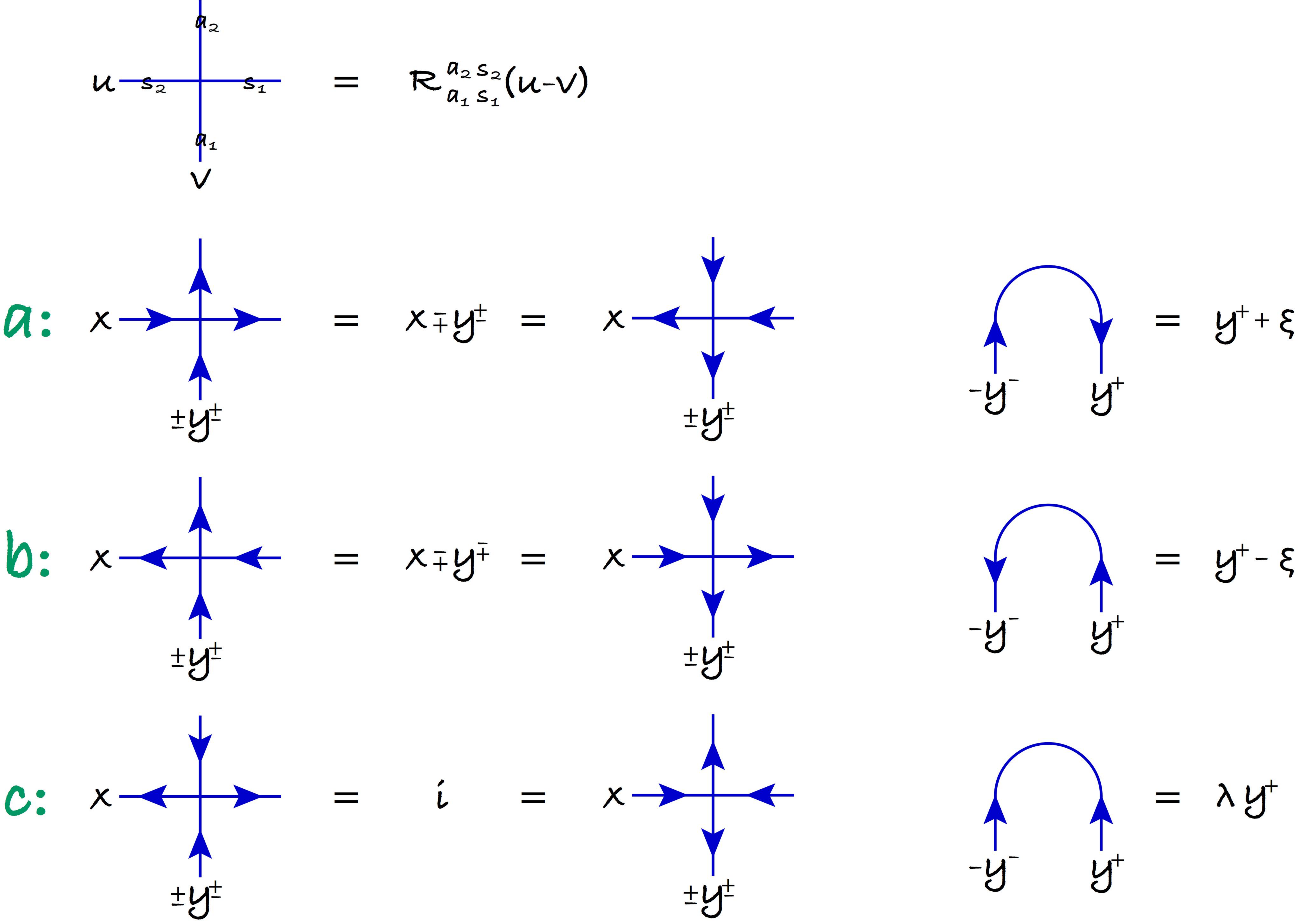}}
\caption{
{\it The left and middle columns show the bulk vertex weights of the rational six-vertex model. 
     The right column shows the boundary vertex weights.}
}
\label{Weights}
\end{center}
\end{figure}

To evaluate the overlaps (\ref{genTsoverlap}) and (\ref{six-vertex-to-Neel}), we proceed along the same 
lines as \cite{Pozsgay:2009,Tsuchiya:qf, Brockmann:2014a}, introducing along the way modifications necessary 
to account for spin-non-preserving term in the boundary state, or equivalently, a non-diagonal term in 
the reflection matrix. The key step is to reformulate the problem in terms of a partition function of 
the rational six-vertex model on a rectangular lattice with a modified, or partial version of Tsuchiya's 
reflecting-boundary domain-wall boundary conditions. It should also be possible to use the recursion 
relation that was derived using the algebraic Bethe ansatz in \cite{Piroli:2014xza}. The Tsuchiya 
determinant is a solution of this recursion relation.
 
%SECTION.03
\section{Partial reflecting-boundary domain-wall partition function}
\label{section.03}

The overlap (\ref{genTsoverlap}) can be represented as a partition function of the rational six-vertex model 
on the $[ M \! \times \! 2N ]$ rectangular lattice, as in Figure {\bf \ref{Zxy}}, where the spin-chain sites 
are associated with the vertical lines and the horizontal lines represent the auxiliary spaces of the 
$B$-operators. The weights of the bulk vertices are the matrix elements of the $R$-matrix, while the weights 
of the boundary vertices are the coefficients of the boundary state (\ref{boundary-state}), as in Figure 
{\bf \ref{Weights}}.
The spins in the bulk are conserved in the sense that each vertex has two in- and two out-bound arrows, as 
in the left and middle columns of Figure {\bf \ref{Weights}}. The spins on the boundary, with reflecting 
boundary conditions at $\lambda=0$ are also conserved, as in the right column of Figure {\bf \ref{Weights}}. 
In this case, spin conservation implies that $M = N$. The partition function on the resulting 
$[ N \! \times \! 2N ]$ lattice admits a determinant representation, as shown by Tsuchiya \cite{Tsuchiya:qf}. 

In this note, we are interested in the more general case where the boundary conditions are partially reflecting, 
and the upper boundary can absorb an excess spin, thus allowing $M$ to be smaller than $N$. The resulting 
statistical mechanical system can be regarded as a degenerate version of Tsuchiya's. The extra horizontal 
lines can be systematically eliminated by taking $(N-M)$ auxiliary-space rapidity variables, in the original 
system, to infinity, and renormalizing the partition function appropriately to obtain a finite result. 
The procedure is described in detail in \cite{Foda:2012yg}, where the partition function of the six-vertex 
model with partial domain-wall boundary conditions was calculated by degenerating the partition function 
of the system with the domain-wall boundary conditions \cite{Korepin:1982gg,Izergin1987}. We do not follow 
this route here. Instead, we follow the derivation of \cite{Tsuchiya:qf}, while taking the more general 
structure of the boundary reflection matrix into account.  

The derivation of \cite{Tsuchiya:qf} relies on a recursion relation which relates the partition functions 
on lattices of different sizes. The recursion relation is derived by f parts of the configurations 
using suitable choices of some of the free variables, and the observation that the type-$a$ or type-$b$ weights 
in Figure {\bf \ref{Weights}} vanish if the vertical and horizontal rapidity variables differ by $\pm \frac{i}{2}$. 
In this note, we generalize this recursion relation to accommodate the partially-reflecting boundary conditions 
at $\lambda \neq 0$. The partition function of the statistical mechanical system that we are interested in is 
completely specified by the following four conditions.

%FIGURE.06
\begin{figure}[t]
\begin{center}
\centerline{\includegraphics[width=9cm]{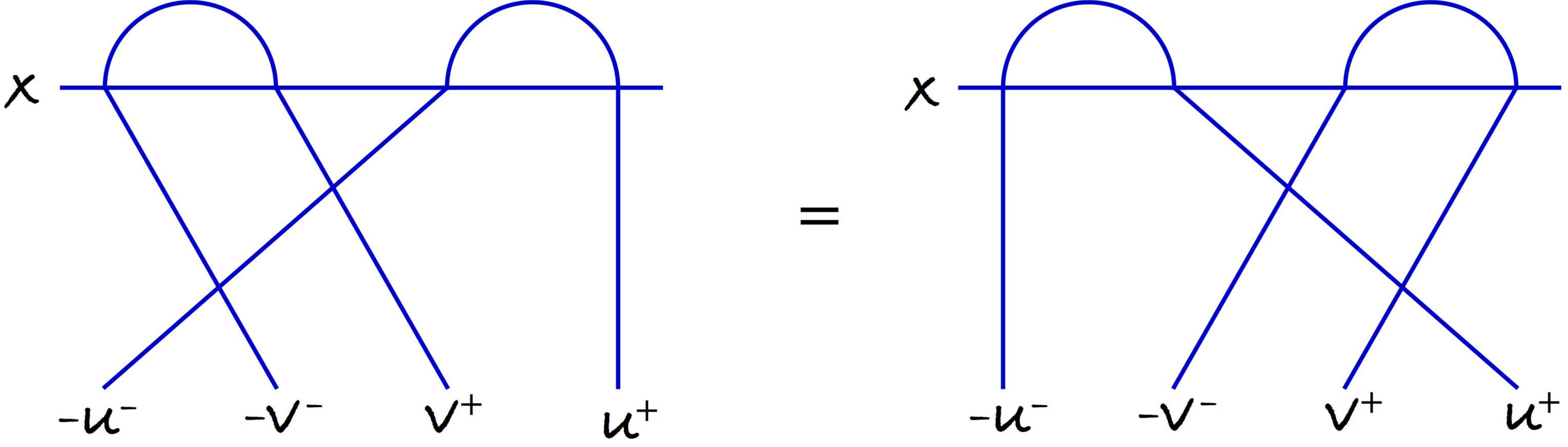}}
\caption{
{\it The reflection equation with one horizontal line added.}
}
\label{R1bar}
\end{center}
\end{figure}

%FIGURE.07
\begin{figure}[t]
\begin{center}
\centerline{\includegraphics[width=9cm]{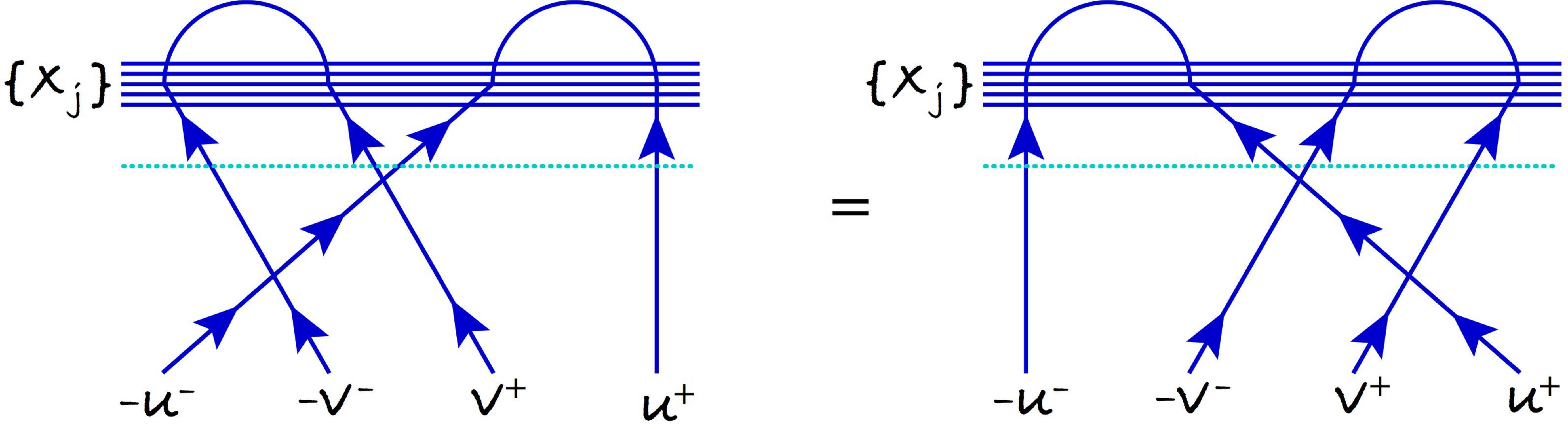}}
\caption{
{\it Once the arrows on the entry lines of the diagram are specified, the spin conservation freezes all 
other spins in the lower part of the diagram, up to the freezing line shown in light blue. The vertices 
below the freezing line are all of $b$-type.}
}
\label{Rfr}
\end{center}
\end{figure}

\subsection{Condition {\bf 1}} 
\label{condition.1}
$Z_{\mathbf{x}\mathbf{y}}$ is symmetric  in $\left\{x_j\right\}$, and separately in $\left\{y_a\right\}$. 
This follows from repeated application of the Yang-Baxter and reflection equations to the partition function. 
The symmetry in $\left\{x_j\right\}$ follows from commutativity of the $B$-operators, which is a consequence 
of the Yang-Baxter equation. The symmetry in $\left\{y_a\right\}$ can be proven by standard manipulations with 
the reflection equation, which we reproduce here for completeness. Multiplying both sides of the reflection 
equation (\ref{reflection-eq}) by 

$$
R_{1a}(x+u^-)R_{3a}(x+v^-)R_{4a}(x-v^+)R_{2a}(x-u^+),
$$
 
\noindent and using the Yang-Baxter equation twice we get the equality depicted in Figure {\bf \ref{R1bar}},

%\begin{eqnarray}
\label{reflection-eq-bis}
%&&\left\langle K_{23}(v)\otimes K_{12}(u)\right|R_{3a}(x+v^-)R_{4a}(x-v^+)R_{1a}(x+u^-)R_{2a}(x-u^+)
%\nonumber \\
%&&\times R_{14}(u+v)R_{13}(u-v)
%\nonumber \\ 
%\nonumber 
%&&=
%\left\langle K_{12}(u)\otimes K_{34}(v)\right|R_{1a}(x+u^-)R_{2a}(x-u^+)R_{3a}(x+v^-)R_{4a}(x-v^+)
%\nonumber \\ 
%\nonumber
%&&\times R_{23}(u+v)R_{24}(u-v).
%\end{eqnarray}

\begin{multline}
\label{reflection-eq-bis}
\\
\left\langle K_{23}(v) \otimes K_{12}(u) \right| R_{3a} (x+v^-) R_{4a}(x-v^+) R_{1a}(x+u^-) R_{2a}(x-u^+) R_{14} (u+v)R_{13}(u-v)
\\ 
=
\left\langle K_{12}(u) \otimes K_{34}(v) \right| R_{1a} (x+u^-) R_{2a} (x-u^+) R_{3a} (x+v^-) R_{4a} (x-v^+) R_{23} (u+v) R_{24} (u-v).
\end{multline}

\noindent The process can be iterated to add an arbitrary number of horizontal lines. The resulting equality is an identity 
of two vectors in $\ll C^2 \rr^{\otimes 4}$.
The next step is to project these vectors on the ground state $\left|\uparrow \uparrow \uparrow \uparrow \right\rangle$, 
in other words to specify all arrows at the bottom of the diagram in Figure {\bf \ref{Rfr}}. By spin conservation, all  
vertices below the freezing line are of type $b$, and consequently produce a common scalar factor 
$R_{\uparrow \uparrow }^{\uparrow \uparrow }(u-v)R_{\uparrow \uparrow }^{\uparrow \uparrow }(u+v)$ on  both sides of 
the equation. The remaining diagrams above the freezing line differ by the order of the vertical rapidity variables $u$ 
and $v$. The same procedure can be applied without much change to swap  any two vertical rapidity variables on the 
$[ M \! \times \! 2N ]$ lattice. 

%FIGURE.08
\begin{figure}[t]
\begin{center}
\subfigure[]{
\includegraphics[height=5.4cm] {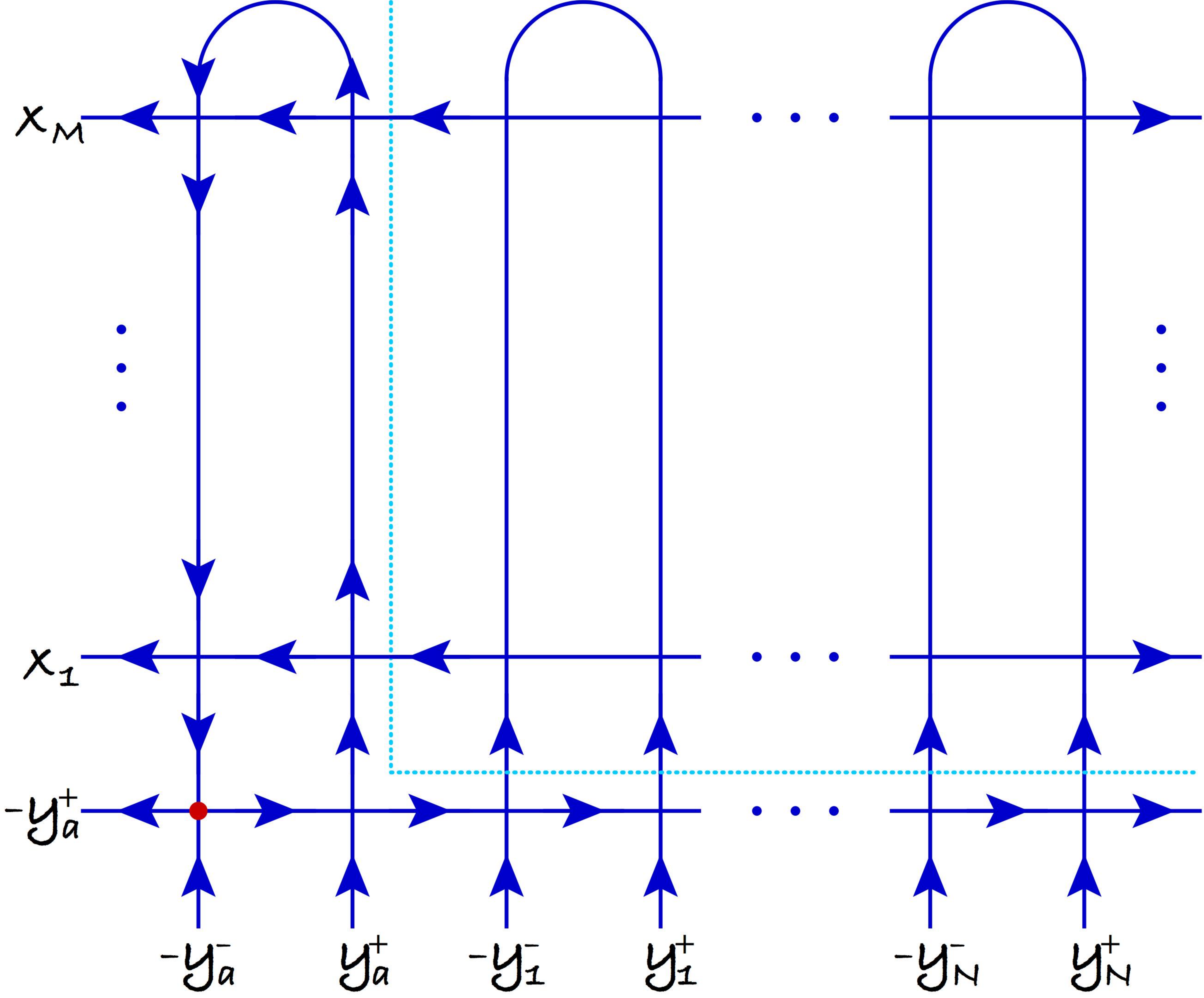}
   \label{fig:subfig1}
}
 \subfigure[]{
\includegraphics[height=5.4cm] {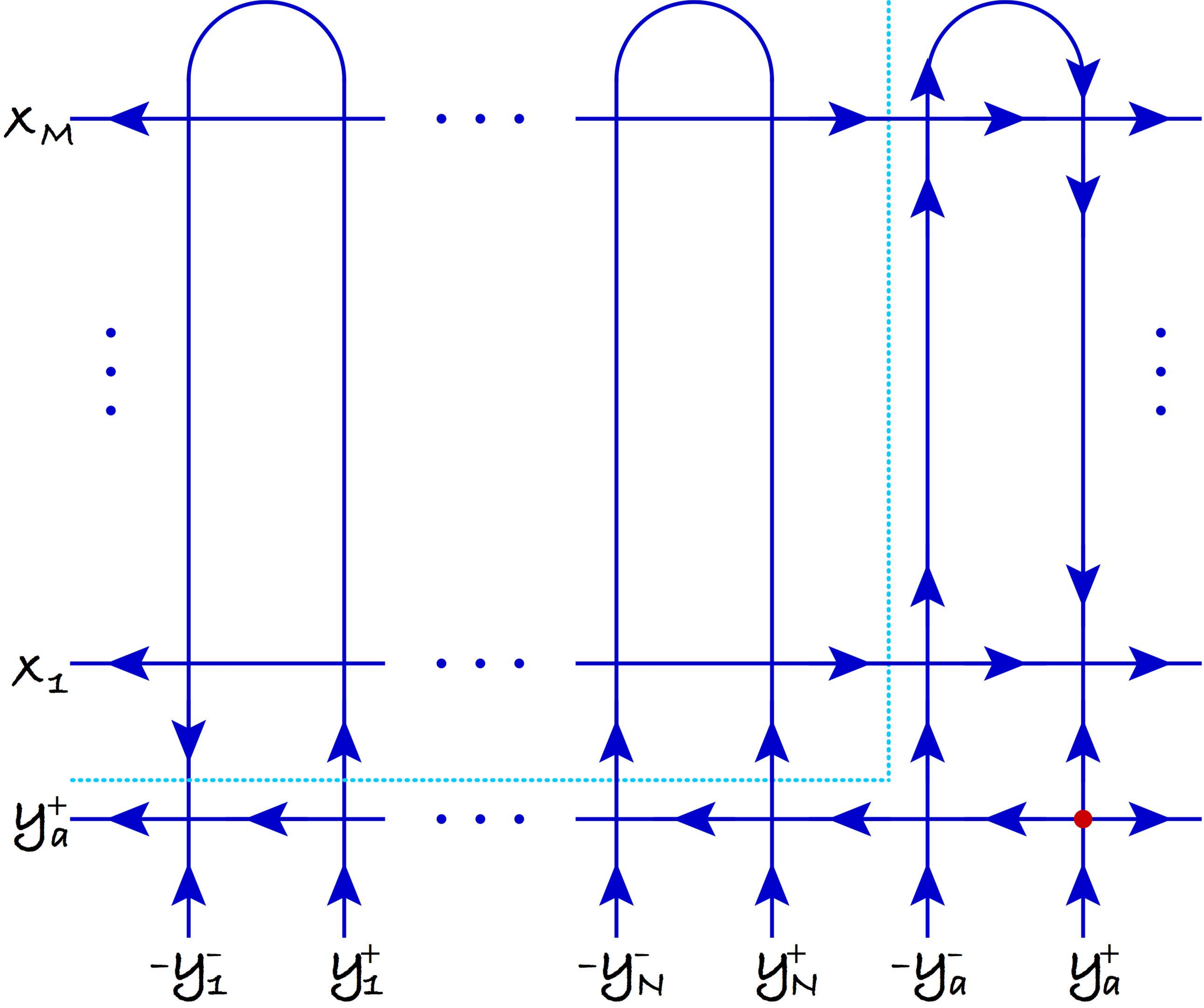}
\label{fig:subfig2}
 }
\caption{
{\it The vertex marked red can only be of the type-$c$ once the condition $x_j=\mp y_a^+$ is imposed. The arrow arrangement 
within the freezing region is fully determined by spin conservation. The freezing trick leads to a recursion relation for 
the partition function.}
}
\label{freetrick}
\end{center}
\end{figure}

\subsection{Condition {\bf 2}} 
\label{condition.2}
$Z_{\mathbf{x}\mathbf{y}}$ is a polynomial of degree $(2N-1)$ in $x_j$. The variable $x_j$ appears in the Boltzmann weights 
associated with the $2N$ vertices on the $j$-th row of the partition function. Each type-$a$ and type-$b$ vertex contributes 
one power of $x_j$, but does not flip the horizontal spin, while a type-$c$ vertex flips the spin but does not depend on $x_j$. 
Since the spin flips at least once, each horizontal line has to contain at least one type-$c$ vertex. Configurations with one 
$c$-vertex and $(2N-1)$ vertices of $a$-type and $b$-type on the same line give a non-zero contribution to the statistical 
ensemble and, consequently, the partition function is a polynomial of degree $(2N-1)$ in $x_j$.

\subsection{Condition {\bf 3}} 
\label{condition.3}
$Z_{\mathbf{x}\mathbf{y}}$ satisfies a recursion relation that equates the partition function on an $[ M \! \times  \! 2N ]$ 
lattice to a partition function on  a smaller $[ (M-1) \! \times \! (2N-2) ]$ lattice, as soon as one of the horizontal 
rapidity variables is set to $x_j=\pm y_a^+$. The recursion relation is derived by the freezing trick illustrated in Figure 
{\bf \ref{freetrick}}. 

First, using the Yang-Baxter and reflection equations, the $j$-th horizontal rapidity and the $a$-th vertical rapidity can 
be moved to the bottom-left corner of the lattice. Spin conservation leaves two possibilities for the bottom-left vertex, 
it is either type-$b$ or type-$c$, as follows from Figure {\bf \ref{Weights}}. If $x_j=-y_a^+$, the type-$b$ vertex has zero 
weight and the bottom-left corner of the partition function is then unambiguously determined. Once the corner weight is fixed, 
the vertices on the two left-most columns and the lower row are recursively determined by spin conservation. This freezes the 
lower and left edges of the partition function as in Figure {\bf \ref{fig:subfig1}}. Similarly, one can freeze the bottom-right 
corner of the partition function by setting $x_j=y_a^+$. The freezing trick results in two recursion relations:

\begin{equation}\label{rec}
\left.Z_{\mathbf{x}\mathbf{y}}\right|_{x_j=\pm y_a^\pm}
=
2 i y_a^\pm
\ll \xi \pm y_a^\pm \rr
\prod_{k \neq j} \ll x_k^2       - \ll y_a^- \rr^2 \rr
\prod_{b \neq a} \ll \ll y_a^{++} \rr^2- y_b^2     \rr
Z_{\hat{\mathbf{x}}_j\hat{\mathbf{y}}_a},
\end{equation}

\noindent where

\begin{equation}
\hat{\mathbf{x}}_j=\left\{x_1, \cdots, \hat{x}_j, \cdots, x_M \right\},
\end{equation}

\noindent and $\hat{x}_j$ means that the variable $x_j$ is omitted.

\subsection{Condition {\bf 4}} At  $M=0$,
\label{condition.4}

\begin{equation}\label{rec0}
 Z_{\emptyset\mathbf{y}}=\lambda ^N\prod_{a}^{}y_a^+.
\end{equation}

\noindent The partition function with no horizontal lines is just a product of $N$ non-reflective boundary 
weights. 

\subsection{Inhomogeneous $\ll N \! \times \! N \rr$ determinant partition function}
\label{inhomogeneous.n.n}
The partition function vanishes for $M>N$. For $M \leq N$, it is completely determined by conditions {\bf 1}, 
{\bf 2}, {\bf 3} and {\bf 4}, for any $M$ and $N$, since a polynomial of degree $(2N-1)$ is completely fixed 
by its values at $2N$ distinct points. Condition {\bf 2}, therefore, determines $Z_{\mathbf{x}\mathbf{y}}$ as 
a function of $x_j$. Eliminating $x$'s one by one, we are left with $Z_{\emptyset\mathbf{y}}$, specified by 
condition {\bf 4}. 

The solution of the recursion relation that satisfies all four conditions can be represented in 
the determinant form

\begin{multline}\label{detTS}
 Z_{\mathbf{x}\mathbf{y}} = i^{N^2+3N+M} \ll \frac{\lambda }{2} \rr^{N-M}
 \prod_j \left( \xi  + x_j \right)
 \prod_a \left( 2y_a + i   \right)
\\
\times  
\frac{
\prod_{ja} \ll \left(x_j-y_a\right)^2+\frac{1}{4} \rr
           \ll \left(x_j+y_a\right)^2+\frac{1}{4 }\rr
}{
\prod_{j<k} \ll x_j^2-x_k^2 \rr
\prod_{a<b} \ll y_a^2-y_b^2 \rr
}
\det \cM,
\end{multline}

\noindent where $\mathcal{M}$ is an $[ N \! \times \! N ]$ matrix,

\begin{equation}\label{matrixM}
\cM_{ab}=
\begin{cases}
y_b^{2a-2}, & \quad a = 1, \cdots, N-M 
\\
\\
\frac{1}{ 
\ll \left(x_j-y_b\right)^2+\frac{1}{4} \rr
\ll \left(x_j+y_b\right)^2+\frac{1}{4} \rr }, & \quad a=N-M+j,~j=1, \cdots, M.
\end{cases}
\end{equation}

Checking the conditions {\bf 1}, {\bf 2}, {\bf 3} and {\bf 4} is straightforward. Symmetry in $x_j$ and $y_a$ is obvious.  
The poles of the prefactor at $x_j=\pm \, x_k$, as well as at $y_a=\pm \, y_b$, are cancelled by the zeros of the determinant. 
Hence, the partition function is a polynomial in each of the $x_j$'s. It is perhaps not immediately obvious why the degree 
of this polynomial is exactly $(2N-1)$, but one can check that the expansion of $\det\mathcal{M}$ at 
$x_j \rightarrow \infty$ starts with $x_j^{-2(N-M)}$, because the lower-order terms are linear combinations of the first 
$(N-M)$ rows of the matrix $\cM$. Checking the recursion relations is also easy, as the $(aj)$ element of $\cM$ develops 
a pole at $x_j = \pm y_a^+$, which eliminates its $(N-M+j)$-th row and $a$-th column.

The determinant representation (\ref{detTS}) generalizes Tsuchiya formula \cite{Tsuchiya:qf}, to which this expression 
reduces when $M=N$. It represents the overlap as an $[ N \! \times \!  N ]$ determinant, where $N$ is half of the length 
of the spin chain. In the sequel, we derive a more compact representation in terms of an $[ M \! \times \! M ]$ determinant, 
where $M$ is the number of magnons, which is general is smaller than $N$. We also study the homogeneous limit when 
we set all the vertical rapidity variables to zero. We should stress that the expression (\ref{detTS}) is valid off-shell, for 
any values of vertical and horizontal rapidity variables. Further, we study the on-shell limit of the partition function when 
the horizontal rapidity variables satisfy the Bethe equations and the state $\left| \mathbf{x} \right\rangle$ is an eigenstate 
of the Heisenberg Hamiltonian.

\subsection{The homogeneous limit of the $\ll N \! \times \! N \rr$ determinant}
\label{homogeneous.n.n}
Observing that the determinant in (\ref{detTS}) scales as $y^{N(N-1)}$, and using 

\begin{multline}
\det_{ab} \ll f_a(y_b^2) \rr \simeq \det_{ab} \ll \sum_{c=1}^N \frac{f^{(c-1)}_a(0)}{(c-1)!} y_b^{2c-2} \rr 
\\
= 
\det_{ac} \ll \frac{f^{(c-1)}_a(0)}{(c-1)!} \rr \det_{db} \ll y_b^{2d-2} \rr 
=
i^{N(N-1)}  \prod_{d<b} \ll y_d^2 - y_b^2 \rr  \det_{ac} \ll \frac{f^{(c-1)}_a(0)}{(c-1)!} \rr,
\end{multline}

\noindent where $\simeq $ means equality up to the leading order in $y^2$, as well as the expansion 

\begin{equation}
 \frac{1}{
 \ll \left(x-y\right)^2+\frac{1}{4}\rr 
 \ll \left(x+y\right)^2+\frac{1}{4}\rr 
 } = \frac{1}{2ix}\sum_{c=1}^{\infty}
 \ll
 \frac{1}{(x^-)^{2c}}-\frac{1}{(x^+)^{2c}}
 \rr 
 y^{2c-2},
\end{equation}

\noindent we find that 

\begin{multline}
\label{off-shell}
\\
 Z_{\mathbf{x}\mathbf{0}}
 = i^{2N^2-N+M^2-M}
 \frac{\lambda ^{N-M}}{2^N}
 \prod_j \ll \frac{x_j+\xi }{x_j} \rr\,\,
 \frac{
 \det_{jk}
 \ll
 (x_j^-)^{2k-2} \, 
 (x_j^+)^{2N  } -
 (x_j^+)^{2k-2} \, 
 (x_j^-)^{2N  }
 \rr
 }{
 \prod_{j<k} \ll x_j^2-x_k^2 \rr
 }\,.
\end{multline}

%SECTION.04
\section{Overlap of a partial N\'eel state and an on-shell Bethe state}
\label{section.04}

To compute the overlap of a partial N\'eel state with a Bethe state, according to (\ref{six-vertex-to-Neel}), 
we 
{\bf 1.} put the inhomogeneous $[ N \! \times \! N ]$ determinant of section {\bf \ref{inhomogeneous.n.n}} 
in a smaller, $[ M \! \times \! M ]$, still inhomogeneous form, 
{\bf 2.} take the homogeneous limit of the $[ M \! \times \! M ]$ determinant, 
{\bf 3.} put the auxiliary-space rapidity variables on-shell by imposing the Bethe equations, then
{\bf 4.} reduce the size of the determinant to $[ \frac{M}{2} \! \times \! \frac{M}{2} ]$. 
The normalized overlap is given by the resulting formula divided by the Gaudin norm of the Bethe vector.

\subsection{An inhomogeneous $\ll M\! \times \! M \rr$ determinant partition function}
\label{inhomogeneous.m.m}

The $[ N\! \times \! N ]$ determinant (\ref{detTS}) admits two, different but equivalent $[ M \! \times \! M ]$ 
representations,

\begin{multline}
\label{alterTS}
Z_{\mathbf{x} \mathbf{y}} =
\left(-2i\right)^M \lambda^{N-M}
\prod_j \ll x_j + \xi \rr
\prod_a y_a^+
\\
\prod_{j<k}
\frac{
\ll \ll x_j^+  \rr^2 - \ll x_k^-    \rr^2 \rr 
\ll \ll x_j^-  \rr^2 - \ll x_k^+    \rr^2 \rr
}{
\ll x_j^2      -  x_k^2 \rr 
\ll \ll x_j^\pm \rr^2 - \ll x_k^\mp \rr^2 \rr
}
\prod_{ja} 
\ll 
\ll x_j^\pm \rr^2 - y_a^2 
\rr
\\
\det_{jk}
\ll
\frac{1}{ \ll x_j^\mp \rr^2 - \ll x_k^\pm \rr^2}
\mp 
\frac{i \delta_{jk}}{2x_j}\,
\prod_a 
\frac{
\ll x_j^\mp \rr^2 - y_a^2
}{
\ll x_j^\pm \rr^2 - y_a^2
}\,
\prod_{l\neq j} 
\frac{
\ll x_j^\pm \rr^2 - \ll x_l^\pm \rr^2
}{
\ll x_j^\mp \rr^2 - \ll x_l^\pm \rr^2}
\rr.
\end{multline}

\noindent The derivation of this result involves standard manipulations of rational sums 
\cite{Izergin1999, Kitanine:2008bs, Kozlowski:2012fv}, the details of which are presented in appendix 
{\bf \ref{ADR}}. In this form, the determinant has almost no dependence on $N$. We reproduce here, for 
completeness, the derivation in \cite{Brockmann:2014a} of the on-shell overlap for the homogeneous spin 
chain, but all the next steps are mathematically the same as in the case of $N=M$ considered in  
\cite{Brockmann:2014a}.

\subsection{The homogeneous limit of the $\ll M \! \times \! M \rr$ determinant}
\label{homogeneous.m.m}

Taking the homogeneous limit in the $[ M \! \times \! M ]$ determinant representation is straightforward 
and yields,

\begin{multline}
\label{homogenB}
\\
\left.Z_{\mathbf{x} \mathbf{0}} \right|_{\lambda =-2i}=
\left(-1\right)^N
\prod_j \ll x_j + \xi \rr \ll x_j^\pm \rr^{2N}
\prod_{j<k}
\frac{
\ll \ll x_j^+ \rr^2 - \ll x_k^- \rr^2     \rr
\ll \ll x_j^- \rr^2 - \ll x_k^+ \rr^2     \rr
}{
\ll     x_j^2         -     x_k^2   \rr 
\ll \ll x_j^\pm \rr^2 - \ll x_k^\mp \rr^2 \rr
}
\det B,
\end{multline}

\noindent where $B$ is an $[ M \! \times \! M ]$ matrix with matrix elements,

\begin{equation}
\label{Bmatrix}
B_{jk}= 
\frac{1}{ 
\ll x_j^\mp \rr^2 - \ll x_k^\pm \rr^2
}
\mp \frac{i \delta_{jk}}{2x_j}
\ll
\frac{x_j^\mp}{x_j^\pm}
\rr^{2N}
\,
\prod_{l\neq j} \frac{ \ll x_j^\pm \rr^2 - \ll x_l^\pm \rr^2}{ \ll x_j^\mp \rr^2 -  \ll x_l^\pm \rr^2}\,.
\end{equation}

\subsection{The overlap of a partial N\'eel state and a generic on-shell Bethe state vanishes}
\label{overlap.vanishes}
Following \cite{deLeeuw:2015hxa}, the MPS (\ref{MPS-state}) is an eigenstate of the degree-3 
conserved-charge operator $\cH_3$, with an $\cH_3$-eigenvalue zero. 
Any on-shell Bethe state, $\chi$, must be an eigenstate of $\cH_3$.
Since the eigenvalue of $\cH_3$ is conserved, the overlap of the MPS with $\chi$, can be non-zero 
only if the $\cH_3$-eigenvalue of $\chi$ is equal to that of the MPS, that is, also zero.
For $\chi$ to have an $\cH_3$-eigenvalue zero, it must be parity-even, that is, the set of 
Bethe roots, $\{ x_j\}$, of $\chi$ must be invariant, as a set, under the parity transformation 
$x_j \rightarrow - x_j$. 
Since the partial N\'eel states are components of the MPS, the same reasoning applies to the overlap 
of a partial N\'eel state and $\chi$
\footnote{\,
Note that the same reasoning does {\it not} apply to the overlap of the MPS and an off-shell Bethe state 
that has an even number of magnons, of the type discussed in subsection \textbf{\ref{homogeneous.m.m}}.
The reason is that the latter is not an eigenstate of $\mathcal{H}_3$, and the $\mathcal{H}_3$ selection rule does not apply to 
its overlap with MPS and consequently with any of the N\'eel states.}. 

Aside from the above reasoning, the technical reason for the vanishing of the overlap, in either 
(\ref{off-shell}) or (\ref{homogenB}), is not immediately obvious. However, it can be proven in 
the latter representation by noting that the matrix $B$ in (\ref{Bmatrix}) has a zero eigenvalue 
if the variables $x_j$ are Bethe roots. This can been seen as follows. Setting the rapidity variables
$\{ x_j\}$ on-shell, the matrix $B$ becomes,

\begin{equation}
B_{jk} = 
\frac{1}{ \ll x_j^\mp \rr^2 - \ll x_k^\pm \rr^2}
\mp 
\frac{i \delta_{jk}}{2x_j}
\prod_{l\neq j} 
\frac{ \ll x_j^\pm \rr^2 - \ll x_l^\pm \rr^2}{ \ll x_j^\pm \rr^2 -  \ll x_l^\mp \rr^2}\,.
\end{equation}

\noindent The fact that this matrix is degenerate follows from the identity

\begin{multline}
\\
\sum_k 
\frac{\pm 2ix_k}{ \ll x_j^\mp \rr^2-  \ll x_k^\pm \rr^2}
\prod_{l\neq k}
\frac{
\ll x_k^\pm \rr^2- \ll x_l^\mp \rr^2}{ \ll x_k^\pm \rr^2 - \ll x_l^\pm \rr^2}
= \oint 
\frac{dz}{2\pi i}
\,\,
\frac{1}{ \ll \ll x_j^\mp \rr^2 - z \rr}
\prod_{l=1}^M 
\frac{ \ll z- \ll x_l^\mp \rr^2 \rr}{\ll z- \ll x_l^\pm \rr^2 \rr}
 =-1,
\label{identity.01}
\end{multline}

\noindent where the contour of integration encircles the poles of the integrand at 
$z= \ll x_l^\pm \rr^2$ counterclockwise, and the last equality is obtained by evaluating 
the residue at infinity. As a consequence of (\ref{identity.01}), the vector with 
components

\begin{equation}
V_k = \pm 2ix_k \prod_{l\neq k} \frac{ \ll x_k^\pm \rr^2- \ll x_l^\mp \rr^2}{ \ll x_k^\pm \rr^2- \ll x_l^\pm \rr^2}
\end{equation}

\noindent can be seen to be a zero eigenvector of $B$,

\begin{equation}
 \sum_{k}^{}B_{jk}V_k=0.
\end{equation}

\noindent 
This implies that a partial N\'eel state has a zero overlap with a generic highest-weight on-shell Bethe state,
where {\it generic} here means a state which is {\it not} invariant under parity transformation, $x_j\rightarrow -x_j$.

%%%%KZ
\subsection{The overlap of a partial N\'eel state and a parity-invariant on-shell Bethe state}
\label{overlap.finite}

If the set of rapidity variables $\{x_j \}$ of the on-shell Bethe state $\chi$ is invariant, as a set, under 
$x_j\rightarrow -x_j$, then $\chi$ is parity-invariant, its $\cH_3$-eigenvalue is zero, and the argument of 
subsection \textbf{\ref{overlap.vanishes}} fails. More concretely, the rapidity variables form pairs of equal 
magnitudes but opposite signs,

\begin{equation}
\label{rapidity.pairs}
\left\{x_j \right\}_{j=1, \cdots, M} = \left\{u_j, -u_j \right\}_{j = 1, \cdots, \frac{M}{2}},
\end{equation}

\noindent which leads to a pole in the prefactor of (\ref{homogenB}). This pole cancels the zero in the determinant 
which lead to the vanishing of the overlap in subsection \textbf{\ref{overlap.vanishes}}. To compute the overlap, 
we resolve the $0/0$ ambiguity, due to the zero and the pole, by shifting the rapidity variables slightly away from 
their parity-invariant values, by defining

\begin{equation}
x_{r j} = r u_j + \varepsilon, \quad r = \pm 1, \quad j = 1, \cdots, \frac{M}{2},
\end{equation}

\noindent calculate the overlap for small but finite $\varepsilon $, then take the limit $\varepsilon \rightarrow 0$.
Some of the matrix elements of $B$ in (\ref{Bmatrix}) diverge as $\varepsilon \rightarrow 0$, but the resulting matrix 
is degenerate and we need to consider also the subleading, $\cO(1)$ term in the determinant. In the following, it is 
convenient to factor out the diagonal matrix $U$, with elements

\begin{equation}
U_{rj, sk} = \frac{r}{2} \, \delta_{r s} \, \delta_{j k} \, u_j^r\,,
\end{equation}

\noindent where $r, s = \pm 1$, and $j, k = 1, \cdots, \frac{M}{2}$. Defining

\begin{equation}
B =  U \hat{B}, 
\end{equation}

\noindent and taking 
into account that $(-u)^\pm =-u^\mp$, we find after some calculation,

\begin{multline}
 \hat{B}_{rj, sk}= 
 \frac{
 \delta_{jk}
 }{
 2 \varepsilon 
 }
 \ll 
 \delta_{rs} + \delta_{r,-s}
 \rr  
 + 2 r u_j^{\mp r} \, \frac{1 - \delta_{jk} \delta_{r,-s}
 }{ 
 \ll u_j^{\mp r} \rr^2 -
 \ll u_k^{\pm s} \rr^2
 }
 \\
 +\delta_{jk} \delta_{rs}
 \ll
 \mp \frac{i u_j^{\mp r}}{u_j} \pm \frac{i N}{u_j^+u_j^-}
 +s \sum_{l \neq j}
 \ll
  \frac{1}{u_j^{\pm r} - u_l^{\mp r}} + \frac{1}{u_j^{\pm r} + u_l^{\pm r}}
 -\frac{1}{u_j-u_l} -\frac{1}{u_j+u_l}
 \rr
 \rr,
\end{multline}

\noindent where we have used the Bethe equations, but only after expanding in $\varepsilon$.
The singular part of the matrix $\hat{B}$ is proportional to the $[1 + \sigma^1]$-projector, 
and has a zero determinant, which why we need to keep the next $\cO(1)$ term. To leading order 
in $\varepsilon$, the matrix $\hat{B}$ has ${M}/{2}$ large eigenvalues, with eigenvectors 
proportional to $[1, \, 1]^t$ and ${M}/{2}$ small eigenvalues, with eigenvectors proportional 
to $[1, \,-1 ]^t$. Denoting the projections on these subspaces as

\begin{equation}
 \hat{B}^{L, S}_{jk} = \frac12 \hat{B}_{rj, sk}
\ll 
\delta^{r s} \pm \delta^{r, - s} 
\rr,
\end{equation}

\noindent where the large and small components are now 
$[ \frac{M}{2} \! \times \! \frac{M}{2} ]$-matrices, and taking into account that 
$\hat{B}^L = \cO (1 / \varepsilon)$ and $\hat{B}^S= \cO (1)$, we have,

\begin{equation}
\det \hat{B} = \det \hat{B}^L \det \hat{B}^S.
\end{equation}

\noindent The large component is 

\begin{equation}
\hat{B}^L_{jk}=\frac{\delta_{jk}}{\varepsilon }\,.
\end{equation}

\noindent To write the small component in compact form, we introduce the following 
notations,

\begin{equation}
K^\pm_{jk}=\frac{2}{\left(u_j-u_k\right)^2+1}\pm\frac{2}{\left(u_j+u_k\right)^2+1}
\end{equation}

\noindent and 

\begin{equation}
G^\pm_{jk} =
K^\pm_{jk}
+
\delta_{jk}
\ll
\frac{2N}{u_j^2 + \frac14} - \sum_l K^+_{jl}
\rr.
\end{equation}

\noindent The small component is 

\begin{equation}
 \hat{B}^S=\pm\frac{i}{2}\,G^+.
\end{equation}

\noindent Hence,

\begin{equation}
\det U = 
\ll-1 \rr^{\frac{M}{2}} 2^{-M} \prod_j \ll \frac{1}{u_j^2 + \frac14} \rr,
\quad 
\det \hat{B^L} = \varepsilon^{- \frac{M}{2}},
\quad 
\det \hat{B}^S = \ll \pm\frac{i}{2} \rr^{\frac{M}{2}}.
\end{equation}

\noindent The $\varepsilon^{-M/2}$ singularity of $\det \hat{B}^L$ cancels the zero in the denominator 
of (\ref{homogenB}), and from (\ref{six-vertex-to-Neel}) and (\ref{homogenB}), we get,

\begin{multline}
\\
 \left\langle \neel_M \right.\!\!\left|\mathbf{u} \right\rangle = 
2 
\ll \frac{i}{2} \rr^M 
\prod_j
\frac{
\ll u_j^2 + \frac14 \rr^{2N+1}
}{
u_j
}\,
\prod_{j<k} 
\frac{
\ll
\left(u_j-u_k\right)^2+1
\rr
\ll
\left(u_j+u_k\right)^2+1
\rr
}{
\ll
u_j^2-u_k^2
\rr^2
}\,
\det G^+.
\end{multline}

Following \cite{Brockmann:2014a}, the determinant expression for the Gaudin norm of a parity-invariant 
on-shell Bethe state \cite{Gaudin:1976sv,Korepin:1982gg}, can be factorized in the form

\begin{equation}
 \left\langle \mathbf{u} \right. \!\! \left|\mathbf{u} \right\rangle =
 \prod_{j}^{}\frac{
\ll u_j^2 + \frac14 \rr^{4N+1}
}{
u_j^2
}\,
 \prod_{j<k}^{}\frac{
\ll
\left(u_j-u_k\right)^2+1
\rr^2
\ll
\left(u_j+u_k\right)^2+1
\rr^2
}{
\ll
u_j^2-u_k^2
\rr^4
}\,
 \det G^+\det G^-.
\end{equation}

\noindent  The normalized overlap is given by
\begin{equation}
\frac{
\left\langle \neel_M \right.\!\!\left|\mathbf{u} \right\rangle
}{
\left\langle \mathbf{u} \right.\!\!\left|\mathbf{u} \right\rangle^\frac12
}
=
2 \ll
\frac{i}{2}
\rr^{\frac{M}{2}}
\ll
\prod_{j}^{}\frac{u_j^2+\frac{1}{4}}{u_j^2}\,\,
\frac{\det G^+}{\det G^-}
\rr^{\frac{1}{2}}.
\end{equation}
This formula was obtained in \cite{Brockmann:2014a} for the N\'eel state with $M=L/2$. The  derivation for arbitrary 
$M$ follows from a symmetry argument and can be found in \cite{Brockmann:2014b,Brockmann:2014c}. Here, we rederive it 
by inspecting the partition function of the six-vertex model with partially reflecting domain-wall boundary conditions.

%SECTION.05
\section{Comments}
\label{section.05}

In \cite{kuperberg2002symmetry}, Kuperberg lists eight classes of domain-wall-type boundary conditions and partition 
functions. These include the original boundary conditions and partition function of Korepin and 
Izergin \cite{Korepin:1982gg,Izergin1987} as well as Tsuchiya's \cite{Tsuchiya:qf}. It is clear that
the remaining six classes admit partial versions in parallel with those discussed in \cite{Foda:2012yg}
and in this note. 

The overlap studied in this note was used in \cite{deLeeuw:2015hxa, Buhl} to compute a class of one-point functions 
in a four-dimensional conformally-invariant supersymmetric gauge theory, in the presence of a defect. The formulation 
of the latter problem in terms of the six-vertex model with particular boundary conditions, may be useful in the sense 
that the six-vertex boundary states may have a direct physical meaning in the gauge theory. 
The boundary state in (\ref{boundary-state}) is a building block of the generalized N\'eel states, and consequently 
of the MPS, which naturally appears in the weak-coupling gauge-theory calculcations \cite{deLeeuw:2015hxa}. Identifying 
a similar building block in the D-brane boundary state that is related, at strong-coupling, to the defect {\it via} the 
AdS/CFT correspondence, would help in finding a fully non-perturbative construction, valid at any coupling.

%SECTION.ACK
\section{Acknowledgements}

O F wishes to thank M Wheeler for collaboration on \cite{Foda:2012yg}. K Z wishes to thank the University of Melbourne 
and the Australian National University, where the work in this note was initiated, for kind hospitality.  
O F is supported by the Australian Research Council. K Z is supported by the Marie Curie network GATIS of the European 
Union's FP7 Programme under REA Grant Agreement No 317089, the ERC advanced grant No 341222, the Swedish Research Council 
(VR) grant 2013-4329, and RFBR grant 15-01-99504. 

%\vfill
%\newpage

\appendix

%SECTION.APP.01
\section{The $\ll M \! \times \! M \rr$ determinant representation}
\label{ADR}

To derive (\ref{alterTS}) from (\ref{detTS}), we introduce two $[ N \! \times \! N ]$ matrices,

\begin{equation}\label{matrixN}
 \cN^\pm_{ba}=
 \frac{
 \prod_l 
 \ll y_b^2 -  \ll x_l^\pm \rr^2 \rr
 }{
 \prod_{c \neq b}
 \ll 
 y_b^2-y_c^2
 \rr
 }
\times 
\begin{cases}
y_b^{2a-2}, & {\rm } a = 1, \cdots, N - M
\\
&
\\          
&  
\\
\frac{
1
}{
y_b^2- \ll x_j^\pm \rr^2
}, & {\rm } a = N - M + j,  \quad j = 1, \cdots, M.
\end{cases}
\end{equation}

\noindent These matrices have the structure similar to (\ref{matrixM}), and while $\mathcal{N}^\pm$ are not 
exactly inverse to $\mathcal{M}$, the product $\mathcal{M}\mathcal{N}^\pm$ is a rather simple matrix with 
a trivial determinant, as we shall see in the moment. We denote the product of $\mathcal{M}$ and 
$\mathcal{N}^\pm$ by $\mathcal{I}^\pm$:

\begin{equation}\label{I=MN}
 \cI^\pm = \cM \, \cN^\pm.
\end{equation}

\noindent The indices of $\cI^\pm_{ad}$ naturally decompose in two sets,  $a =1, \cdots, N-M$ and $a=N-M+j$ 
with $j= 1, \cdots, M$, as in (\ref{matrixM}) and (\ref{matrixN}). The matrix $\cI^\pm$  therefore consists 
of four blocks, $\cI^\pm_{ad}$, $\cI^\pm_{ak}$, $\cI^\pm_{jd}$, and $\cI^\pm_{jk}$, where the indices take 
values $a,d=1,\ldots ,N-M$ and  $j,k=1,\ldots, M$. We use the shorthand notation for the $(N-M+j)$-th index 
of $\cI^\pm$ by simply omitting $N-M$ in the label. The key observation is that $\mathcal{I}^\pm_{ak}$ is 
zero. Indeed, from (\ref{matrixM}), (\ref{matrixN}),

\begin{equation}
\cI^\pm_{ak}=\sum_{b} y_b^{2a-2} \,
\frac{\prod_{l \neq k} \ll y_b^2 - \ll x_l^\pm \rr^2 \rr}{\prod_{c\neq b}
\ll y_b^2-y_c^2 \rr}
=
\oint\frac{dz}{2\pi i}\,\,z^{a-1}\,
\frac{\prod_{l\neq k} \ll z- \ll x_l^\pm \rr^2 \rr}{\prod_c \ll z - y_c^2 \rr}\,,
\end{equation}

\noindent where the contour of integration encircles the set of points $\{y^2_a\}$ counterclockwise. 
But the integrand has no singularities outside the contour of integration. In particular the pole infinity 
vanishes because the integral behaves as $z^{a+M-N-2}$, and $2+N-M-a$ is always bigger than one. Therefore, 
$\cI^\pm_{ak}=0$ and consequently the matrix $\mathcal{I}^\pm$, in the block form, is lower triangular,

\begin{equation}
\cI^\pm=
\ll
\begin{array}{ccc}
\cI^\pm_{ad} &   &  0            \\ 
             &   &               \\
\cI^\pm_{jd} &   &  \cI^\pm_{jk} \\ 
 \end{array}
 \rr.
\end{equation}

The other, non-zero components of $\mathcal{I}^\pm$ can be computed by the same trick. For the sake of 
calculating the determinant of $\mathcal{I}^\pm$, we only need its block-diagonal components, for which 
we have,

\begin{multline}
\cI^\pm_{jk} = 
\sum_b 
\frac{1
}{
\ll y_b^2 - \ll x_j^+ \rr^2 \rr \ll y_b^2 - \ll x_j^- \rr^2 \rr}\,\,
\frac{\prod_{l\neq k} \ll y_b^2- \ll x_l^\pm \rr^2 \rr}{\prod_{c\neq b}
\ll y_b^2-y_c^2 \rr}
\\
=
\oint \frac{dz}{2\pi i}
\frac{1}{
\ll z - \ll x_j^+ \rr^2 \rr 
\ll z - \ll x_j^- \rr^2 \rr}
\frac{\prod_{l\neq k} \ll z- \ll x_l^\pm \rr^2 \rr
}{
\prod_c
\ll z-y_c^2 \rr}
\\
= \pm \frac{1}{2ix_j}
\ll
 \frac{\prod_{l\neq k} \ll \ll x_j^\mp \rr^2- \ll x_l^\pm \rr^2 \rr}{\prod_c \ll \ll x_j^\mp \rr^2 - y_c^2 \rr} -\delta_{jk}\,
 \frac{\prod_{l\neq j} \ll \ll x_j^\pm \rr^2- \ll x_l^\pm \rr^2 \rr}{\prod_c \ll \ll x_j^\pm \rr^2 - y_c^2 \rr}
\rr,
\end{multline}

\noindent where the last equality is obtained by inflating the contour of integration and computing the residues at 
$z= (x_j^\mp)^2$ and $z=(x_j^\pm)^2$. The latter residue vanishes unless $j=k$.
For the $ad$ components, we get

\begin{equation}
 \cI^\pm_{ad} =
 \sum_b y_b^{2a+2d-4}\,
 \frac{
 \prod_l \ll y_b^2 - \ll x_l^\pm \rr^2 \rr
 }{
 \prod_{c \neq b}
 \ll y_b^2 -y_c^2 \rr
 }
 =
 -\mathop{\mathrm{res}}_{z=\infty }
 z^{a+d-2}\, 
 \frac{
 \prod_l \ll z - \ll x_l^\pm \rr^2 \rr
 }{
 \prod_c \ll z -  y_c^2      \rr
 }\,.
\end{equation}

\noindent The residue on the right-hand-side vanishes for $a+d<N-M$ and equals one for $a+d=N-M$. 
The $[ (N-M) \! \times  \! (N-M) ]$ matrix with elements $\cI^\pm_{ad}$ therefore has 
a triangular form

\begin{equation}
 \cI^\pm_{ad}=
\ll
 \begin{array}{ccc}
 0  &  &  1 \\ 
   & \cdots &  \\ 
 1  &  & * \\ 
 \end{array}
\rr,
\quad
\det_{ad} \cI^\pm_{ad}=\left(-1\right)^{\frac{\left(N-M-1\right)\left(N-M\right)}{2}}.
\end{equation}

As a consequence of (\ref{I=MN}),

\begin{equation}
 \det \cM =\frac{\det \cI^\pm}{\det \cN^\pm}
 =\left(-1\right)^{\frac{\left(N-M-1\right)\left(N-M\right)}{2}}\frac{\det_{jk}
 \cI^\pm_{jk}}{\det \cN^\pm}\,.
\end{equation}

\noindent The denominator in this formula is a generalized Cauchy determinant that can be explicitly 
calculated

\begin{equation}
 \det \cN^\pm=
 \left( -1 \right)^{\left(N+1\right)M}
 \,
 \frac{
 \prod_{j<k}
 \ll \ll x_j^\pm \rr^2- \ll x_k^\pm \rr^2 \rr
 }{
 \prod_{a<b} \ll y_a^2-y_b^2 \rr
 }\,.
\end{equation}

\noindent Collecting the pieces we get,

\begin{multline}
\det \cM =
\left(-1\right)^{\frac{N\left(N-1\right)}{2}+M}
\\
\frac{\prod_{a<b}  \ll y_a^2-y_b^2 \rr}{\prod_{ja} \ll \ll x_j^\mp \rr^2 - y_a^2 \rr}
\prod_{j<k} \frac{ \ll \ll x_j^+ \rr^2- \ll x_k^- \rr^2 \rr
                   \ll \ll x_j^- \rr^2- \ll x_k^+ \rr^2 \rr
                   }{
		               \ll x_j^\pm \rr^2 - \ll x_k^\pm \rr^2}
\\
\times 
 \det_{jk}
 \ll
 \frac{1}{ \ll x_j^\mp \rr^2 - \ll x_k^\pm \rr^2}
 \mp \frac{ i \delta_{jk}}{2x_j}\,
 \prod_a \frac{ \ll x_j^\mp \rr^2 - y_a^2}{ \ll x_j^\pm \rr^2 - y_a^2}\,
 \prod_{l\neq j} \frac{ \ll x_j^\pm \rr^2 - \ll x_l^\pm \rr^2}{ \ll x_j^\mp \rr^2 - \ll x_l^\pm \rr^2}
 \rr.
\end{multline}

\noindent Equation (\ref{alterTS}) then follows from (\ref{detTS}).

\vfill
\newpage

\end{document}